\documentclass[aps,twocolumn,superscriptaddress,longbibliography, floatfix,pra]{revtex4-1}

\usepackage{lipsum}
\usepackage{graphicx}
\usepackage{amsmath,amssymb}
\usepackage{braket}
\usepackage{mathtools}
\usepackage{hyperref}
\usepackage{multirow}
\usepackage{color}
\usepackage[normalem]{ulem}
\usepackage{amsfonts}
\usepackage{float}
\usepackage{pdfpages} % include appendix
\makeatletter
\usepackage{subfigure}

\def\beq{\begin{equation}}
\def\eeq{\end{equation}}
\def\bea{\begin{eqnarray}}
\def\eea{\end{eqnarray}}

\AtBeginDocument{\let\LS@rot\@undefined}
\makeatother

\begin{document}

\title{Distinct universality classes of diffusive transport from full counting statistics}

%\title{Theory of anomalous full counting statistics in anisotropic spin chains}

\author{Sarang Gopalakrishnan}
\affiliation{Department of Physics, The Pennsylvania State University, University Park, PA 16802, USA}

\author{Alan Morningstar}
\affiliation{Department of Physics, Princeton University, Princeton, NJ 08544, USA}
\author{Romain Vasseur}
\affiliation{Department of Physics, University of Massachusetts, Amherst, MA 01003, USA}
\author{Vedika Khemani}
\affiliation{Department of Physics, Stanford University, Stanford, CA 94305, USA}

\begin{abstract}

The hydrodynamic transport of local conserved densities furnishes an effective coarse-grained description of the dynamics of a many-body quantum system. However, the full quantum dynamics contains much more structure beyond the simplified hydrodynamic description. Here we show that systems with the same hydrodynamics can nevertheless belong to distinct dynamical universality classes, as revealed by new classes of experimental observables accessible in synthetic quantum systems, which can, for instance, measure simultaneous site-resolved \emph{snapshots} of all of the particles in a system.  Specifically, we study the full counting statistics of spin transport, whose first moment is related to linear-response transport, but the higher moments go beyond. We present an analytic theory of the full counting statistics of spin transport in various integrable and non-integrable anisotropic one-dimensional spin models, including the XXZ spin chain. We find that spin transport, while diffusive on average, is governed by a distinct non-Gaussian dynamical universality class in the models considered.    
We consider a setup in which the left and right half of the chain are initially created at different magnetization densities, and consider the probability distribution of the magnetization transferred between the two half-chains.
We derive a closed-form expression for the probability distribution of the magnetization transfer, in terms of random walks on the half-line. 
We show that this distribution strongly violates the large-deviation form expected for diffusive chaotic systems, and explain the physical origin of this violation. We discuss the crossovers that occur as the initial state is brought closer to global equilibrium. Our predictions can directly be tested in experiments using quantum gas microscopes or superconducting qubit arrays.

\end{abstract}
\vspace{1cm}

\maketitle

\section{Introduction}

The dynamics of quantum systems is traditionally characterized by equilibrium spatiotemporal correlation functions of local observables, which probe the transport of conserved quantities and response functions of local observables, and shed light on the universal properties of near-equilibrium (and generally low temperature) quantum condensed matter. The computation of dynamical correlators is generally intractable in a many-body system; however, in generic systems, local operators ``thermalize''---i.e., their expectation values approach those in the appropriate equilibrium state~\cite{rigolreview}. Assuming thermalization occurs locally in a spatially extended system, the coarse-grained evolution of slow local expectation values follows from hydrodynamics. Thus, in quantum systems, hydrodynamics is an emergent property of the expectation values of local operators. 

Breakthrough experimental advances in building quantum simulators provide novel probes into the dynamics of quantum systems that go well beyond measurements of local expectation values. For example, experiments in platforms such as ultracold atoms or superconducting qubits ~\cite{gross2017quantum, arute2019quantum, scholl2021microwave} can take simultaneous \emph{snapshots} of all the particles in a system, and can access dynamics in highly excited states, at energy densities corresponding to infinite temperature. These advances have motivated intense interest in understanding dynamical universality classes of quantum matter that is strongly out of equilibrium and far from low-temperature, linear-response regimes. In this regard, much recent effort has focused on systems that fully fail to reach thermal equilibrium, i.e. systems in which transport is arrested and hydrodynamics breaks down~\cite{RevModPhys.91.021001}. 

In this work, we ask a different question: can quantum systems that follow the \emph{same} hydrodynamics, given (e.g.) by the usual diffusion equation, nevertheless belong to distinct dynamical universality classes?  We answer this question in the positive, and show analytically that systems with the same hydrodynamics can belong to qualitatively distinct universality classes distinguished by the behavior of more detailed observables such as the statistics of full-system snapshots. Our work illustrates the wealth of interesting phenomena that remains to be discovered even in seemingly well-understood dynamical regimes (high temperature and diffusive), and the central role played by novel quantum simulation experiments in exposing these phenomena.   

A quantity that compactly encapsulates the statistics of snapshots is the ``full counting statistics'' (FCS) of conserved charges~\cite{levitov1993charge,doi:10.1063/1.531672,PhysRevB.51.4079,PhysRevB.67.085316,PhysRevLett.110.050601, TOUCHETTE20091, garrahan2018aspects, Essler_xxz_fcs, Essler_Ising_fcs}. We consider the following setup: First, initialize a one-dimensional lattice with both the left and right half-systems in definite charge states. Then, after evolving for a time $t$, measure the amount of charge that has been transferred from the left to the right  half-system. Repeating this experiment many times yields a \emph{quantum} distribution of measurement outcomes. The first cumulant of the FCS is related to linear-response transport, but the higher cumulants go beyond it (we emphasize that these are cumulants over a measurement ensemble, \emph{not} spatial cumulants of the dynamic structure factor). 

For classical systems, one can ask an analogous question about the distribution over noise realizations and/or initial thermal ensembles: in stochastic models such as the exclusion process~\cite{Derrida:2009aa,PhysRevLett.107.010602,FCSDerridaASEP,PhysRevLett.109.170601}, or growth processes governed by the Kardar-Parisi-Zhang (KPZ) equation~\cite{kpz,KPZFCS}, this distribution is known exactly, while tensor-network methods can be used to efficiently compute it more generally~\cite{PhysRevLett.123.200601}. 
A general property of these distributions is that they obey a large-deviation principle~\cite{TOUCHETTE20091,Lazarescu_2015}: for chaotic diffusive systems, all cumulants scale the same way with time.
For deterministic classical systems, surprisingly, this general expectation can fail: this failure was recently discovered in studies of classical integrable spin chains and cellular automata~\cite{PhysRevLett.128.090604, krajnik2022exact}.

For \emph{quantum} distributions, much less is known beyond free fermions and systems with ballistic transport~\cite{10.21468/SciPostPhys.8.1.007,DoyonFCS,10.21468/SciPostPhys.10.5.116,doi:10.1073/pnas.2106945118, krajnik2022exact}. 
The question of FCS in interacting quantum systems was recently revived because of the surprising discovery that spin transport in the one-dimensional Heisenberg model appears to be described by the KPZ equation~\cite{idmp,lzp,PhysRevLett.122.127202,PhysRevLett.122.210602,gvw,PhysRevLett.123.186601,PhysRevLett.125.070601,vir2019,Scheie2021,2107.00038,PhysRevX.11.031023,Bulchandani_2021}, although the Heisenberg model is an integrable quantum system with no stochasticity. In Ref.~\cite{2107.00038} the FCS was experimentally extracted using a quantum gas microscope, and its first three cumulants were found to agree with KPZ predictions. Nevertheless, many basic questions about the FCS of quantum systems remain open.

Here we address this question for the spin chains with easy-axis anisotropy, focusing primarily on the XXZ spin chain.  
Introducing an easy-axis anisotropy makes linear-response transport diffusive~\cite{dbd2,PhysRevLett.122.127202}, suggesting that the FCS might follow that of diffusive classical systems such as the symmetric exclusion process. 
Instead, we find that it is strongly anomalous, with the mean and variance of the FCS scaling as different powers of time. 
We explain the physical origin of this anomalous FCS and determine an exact functional form, including prefactors, for the FCS at late times, which we argue holds everywhere in the easy-axis regime of the XXZ spin chain. 
As we will discuss, this anomalous FCS is related to a breakdown of the central limit theorem due to strong correlations between successive scattering events~\footnote{This point was also previously made in Ref.~\cite{krajnik2022exact}.}. 
Our results generalize to nonintegrable systems with hardcore kinetic constraints; there, they give \emph{subdiffusion}~\cite{PhysRevLett.127.230602,2021arXiv210913251D} with the same universal FCS. 

The rest of this paper is organized as follows. In Sec.~\ref{setup} we specify the models, observables, and initial conditions of interest. In Sec.~\ref{lowfilling} we analyze a limit (a low-entropy limit where the left/right halves are almost maximally polarized) in which the FCS can be computed by elementary means. In Sec.~\ref{general_filling} we generalize this result to arbitrary filling and discuss the crossover between equilibrium and nonequilibrium initial states. In Sec.~\ref{numerics} we present numerical data on FCS in quantum spin chains that is consistent with our theoretical predictions. We show that our main results are not restricted to integrable systems, but also extend to constrained systems in the presence of noise. Finally in Sec.~\ref{conc} we summarize our results and comment on open questions.

\section{Setup and background}\label{setup}

In the course of this paper we will consider a family of closely related models, which we enumerate here for clarity. All the models on this list exhibit the same transport phenomena, with the same universal FCS, but each has some practical advantages:

\begin{enumerate}
\item \emph{XXZ spin chain}. The canonical model to which our results apply is the easy-axis ($\Delta > 1$) regime of the XXZ quantum spin-$\frac{1}{2}$ chain 
\beq\label{XXZ}
H =  -\sum\nolimits_i S^x_i S^x_{i+1} + S^y_i S^y_{i+1} + \Delta S^z_i S^z_{i+1}.
\eeq
When writing down analytic expressions for prefactors we will focus on this model, since a great deal is known quantitatively about its quasiparticle structure.
\item \emph{Trotterized (Floquet) XXZ}. To study the dynamics of the XXZ model numerically one can discretize its time evolution via a Trotter decomposition. For the naive Trotterization of the XXZ model, the small time steps required to converge the time evolution are numerically expensive. To faithfully study XXZ dynamics while keeping the time-step relatively large, we rely on the integrable Trotterization in Ref.~\cite{PhysRevLett.121.030606, Ljubotina-Prosen2019_ballistic}. This is a periodic discrete-time (Floquet) integrable quantum model that has the same hydrodynamic behavior as the XXZ spin chain for arbitrary time step. 

\item \emph{Folded XXZ automaton}. Trotterizing the XXZ model yields something more amenable to numerical study; however, quantum time evolution is still prohibitive for system sizes above $L\agt 32$ spins. A more direct test of our predictions comes from a classically simulable limit of the model, which we now describe. In the limit $\Delta \to \infty$, the XXZ spin chain becomes a model with constrained hopping, 
\beq\label{foldedxxz}
H_{\mathrm{folded}} = - \sum\nolimits_i \Pi_{i-1, i+2} (S^x_i S^x_{i+1} + S^y_i S^y_{i+1}),
\eeq
where the projector $\Pi_{i-1,i+2} \equiv (1 + 4S^z_{i-1} S^z_{i+2})/2$ enforces conservation of number of domain walls, as appropriate to the $\Delta \rightarrow \infty$ limit. The folded XXZ model was initially proposed as a Hamiltonian~\cite{10.21468/SciPostPhysCore.4.2.010}; however, each term in it is both unitary and ``automaton-like'' in the sense that it maps a computational-basis product state to another computational-basis product state. The resulting automaton was shown to be integrable in Ref.~\cite{balazs1} if the gates are applied in a particular pattern. As an automaton, its dynamics starting from a computational-basis product state can be classically simulated, allowing us to go to late times and large systems. 

\item\emph{XNOR model}. Finally, we will consider a \emph{nonintegrable} constrained stochastic model in which each term in Eq.~\eqref{foldedxxz} is applied at random~\cite{PhysRevLett.127.230602,2021arXiv210913251D}. Remarkably, this model also exhibits the same FCS as the integrable models specified above, with a different dynamical exponent $z=4$ instead of $z=2$.

\end{enumerate}

All the models listed above conserve the total magnetization $\sum\nolimits_i S^z_i$. We will consider the transport of magnetization (i.e., of spin). 
At half-filling (zero field) and any finite temperature, linear-response spin transport is known to be diffusive, even though energy transport is ballistic. In order to characterize non-linear,  far from equilibrium,   transport properties, we focus here on a domain wall (DW) initial state with an initial charge (spin) imbalance~\cite{PhysRevE.81.061134,gamayun2019domain,PhysRevB.97.081111,Misguich2019}. We prepare the system at time $t=0$ in a high temperature ($T=\infty$) pure state in which the left (right) half of the system has average magnetization $\langle S^z \rangle = +m/2$ ($-m/2$). For specificity we take the left and right half-systems to be separately drawn from the \emph{canonical} ensemble, so for each run of the experiment the initial state has a definite particle number in each half-chain. For $m$ small, the system is near equilibrium and we expect to recover linear-response results, while $m=1$ corresponds to a fully polarized pure initial state $\left| \uparrow \uparrow \dots \uparrow \downarrow \dots \downarrow \downarrow \right. \rangle $. We consider periodic boundary conditions (PBC), and let the system evolve unitarily following the Hamiltonian~\eqref{XXZ}. 
Our goal is to characterize the statistics of the magnetization transferred between the two half chains as a function of time
\beq\label{eqP}
{\hat Q}(t) = \sum_{i \geq 0} [S_i^z(t) - S_i^z(0)] - \sum_{i<0} [S_i^z(t) - S_i^z(0)].
\eeq
For any initial magnetization imbalance $m<1$, the average magnetization transfer obeys a dynamical scaling $\langle {\hat Q}(t) \rangle  \sim t^{1/z}$,
with $z=2$, characteristic of a diffusive system. In what follows, we will be interested in the full statistics of this magnetization transfer $Q(t)$, that we will characterize using the full counting statistics (FCS) function 
\beq\label{eqFCS}
\chi_t(\lambda) = \log \langle {\rm e}^{i \lambda {\hat Q}(t)} \rangle \equiv \log \int d  Q P_t( Q) {\rm e}^{i \lambda  Q},
\eeq
which is the generating function of the cumulants of ${\hat Q}$. Equivalently we aim to characterize the full distribution function $P_t( Q)$ defined in Eq.~\eqref{eqFCS}, where $ Q$ are measurement outcomes of ${\hat Q}$.

\begin{figure}[tb]
\begin{center}
\includegraphics[width = 0.45\textwidth]{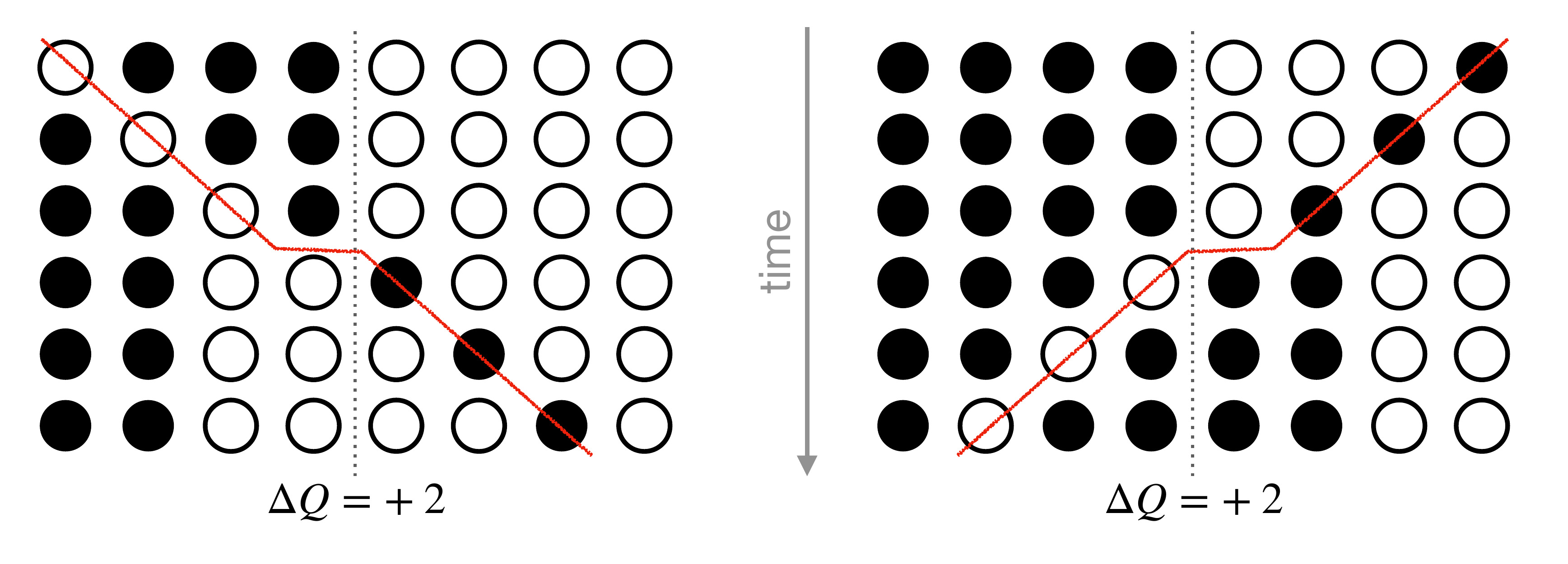}
\caption{Illustrative trajectories in which a magnon goes through the domain wall starting either from the left or the right. The magnon trajectory is indicated in red; the ``cut'' dividing the left and right half-chains is shown as a gray dashed line. The passage of the magnon shifts the domain wall by two steps. Whichever way the domain wall shifts, provided it moves away from the cut, one unit of magnetization (i.e., one net black circle) moves from the left half-chain to the right half-chain. According to our conventions (Eq.~\eqref{eqP}) this gives $\Delta Q = +2$. If we run this process in reverse, the domain wall moves toward the cut and magnetization is transferred to the left.}
\label{fig1}
\end{center}
\end{figure}

For a typical many-body diffusive system, classical or quantum, we expect the fluctuations of $Q(t)$ to be controlled by the variance scaling as $\langle {\hat Q}(t)^2 \rangle_c \equiv \langle{\hat Q}(t)^2 \rangle- \langle {\hat Q}(t) \rangle^2 \sim \sqrt{t}$. This can be seen easily for a classical system of independent random walkers, for which the variance is simply controlled by the central limit theorem, so that $ \sqrt{\langle {\hat Q}(t)^2 \rangle_c}/\langle {\hat Q}(t) \rangle \sim 1/\sqrt{\langle {\hat Q}(t) \rangle } \sim 1/t^{1/4}$. This behavior generalizes to other diffusive interacting systems~\cite{10.1143/PTPS.184.276}, such as the symmetric exclusion process for example~\cite{Derrida:2009aa}, and in general, the FCS in a diffusive system is expected to scale as $\chi_t(\lambda)\sim \sqrt{t} F(\lambda) $, so that all cumulants of the magnetization (``charge'') transfer scale as $\sqrt{t}$. Equivalently, the distribution of the magnetization transfer obeys a large deviation principle
\beq \label{eqLargedeviation}
 P_t \left(\frac{ Q}{\sqrt{t}}= q \right) \sim {\rm e}^{\sqrt{t} G(q)},
\eeq
where the so-called large deviation function $G(q)$ can be related to $F(\lambda)$ by a Legendre transform. 

Although the XXZ spin chain~\eqref{XXZ} also has diffusive magnetization transfer on average, we will argue below that the expected large-deviation  form~\eqref{eqLargedeviation} breaks down. In particular, the rescaled variance $\langle {\hat Q}(t)^2 \rangle_c /\sqrt{t}$ {\em diverges} as a function of time, and we will show that the width is of the same order as the average: $\sqrt{\langle {\hat Q}(t)^2 \rangle_c} \sim \langle {\hat Q}(t) \rangle  \sim \sqrt{t}$. The origin of this anomalous behavior is the unusual physical mechanism leading to spin diffusion in the XXZ spin chain. In what follows, we will focus on the asymptotic, long-time behavior of the distribution $P_t( Q)$, which will describe fluctuations over scale $Q \sim \sqrt{t}$ for $m>0$, and $Q \sim t^{1/4}$ in equilibrium $m=0$.

\begin{figure*}[!t]
\begin{center}
\includegraphics[width = 0.98\textwidth]{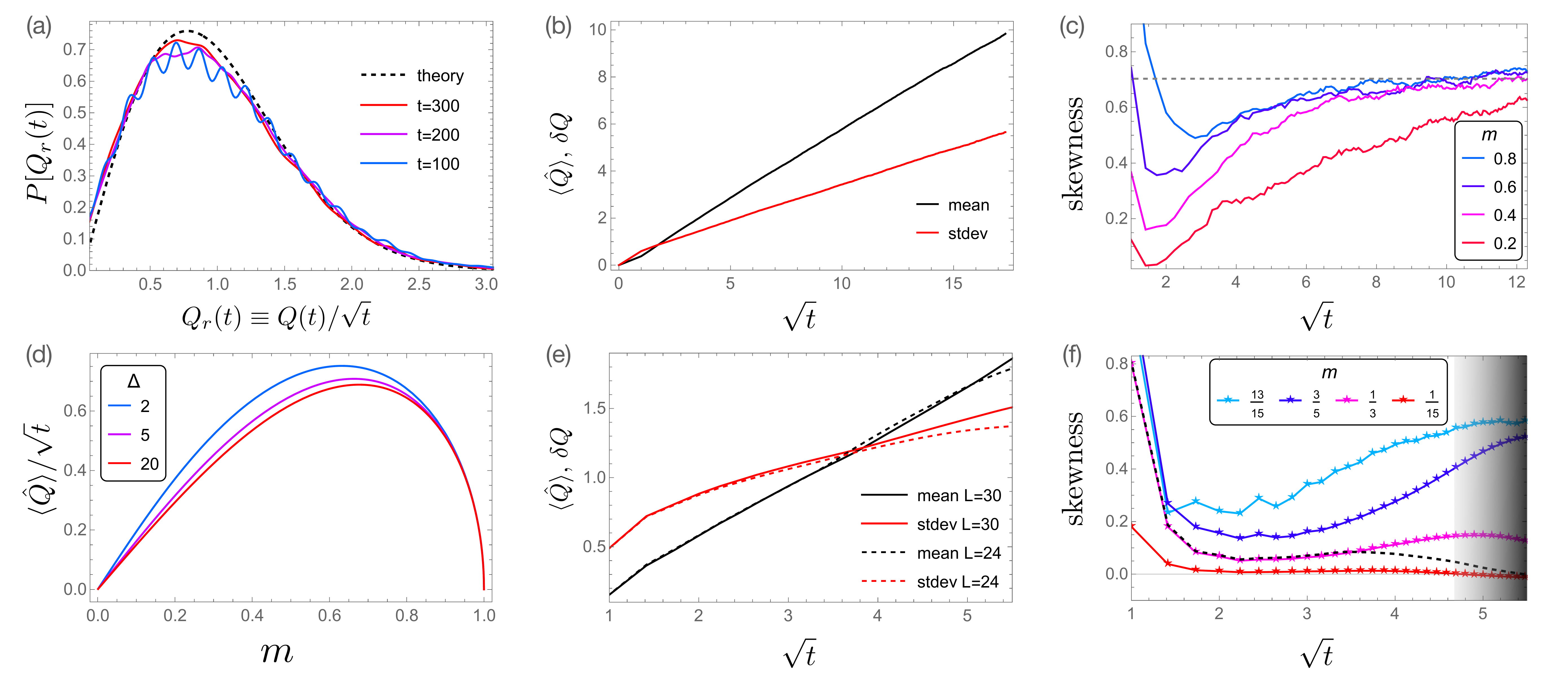}
\caption{ 
Upper panels: results on the automaton~\cite{balazs1}. All results are averaged over between $3\times 10^4$ and $10^5$ initial states. (a)~Distribution of the magnetization transfer $Q(t)$ for $m = 0.6$ at various times, rescaled by the mean transfer $\sqrt{t}$. These collapse well onto each other and onto the theoretical prediction~\eqref{eqDistribution}. (b)~Evolution of the mean and standard deviation of $Q(t)$ at $m = 0.6$, indicating that both go as $\sqrt{t}$. (c)~Evolution of the skewness of $Q(t)$ for various values of initial magnetization $m$; as $m \to 0$ the skewness takes longer to grow to its universal late-time value. All simulations were done on system sizes $L = 12 t_{\mathrm{max}}$ where $t_{\mathrm{max}}$ is the latest time presented. For this size, finite-size effects are demonstrably absent. Lower panels: results on the XXZ spin chain and its Trotterization. (d)~Theoretically predicted coefficient of mean magnetization transfer as a function of anisotropy $\Delta$ and magnetization $m$. Note that this goes to zero both at $m = 0$ (by reflection symmetry) and at $m = 1$ (because the domain wall freezes). (e)~Mean and standard deviation of $Q(t)$ at $m = 1/3$ from numerical simulations of the Trotterized XXZ spin chain (see main text). Solid (dashed) lines are for $L = 30$ ($L = 24$) and their separation indicates where finite size effects manifest themselves for $L = 24$. (f)~Skewness for various values of $m$ at $L = 30$. The black dashed line is for $L = 24$; from it, we can infer that finite-size effects cause a downturn of the skewness, which appears to set in for $L = 30$ in the shaded region at the right end of the plot.}
\label{balazsfig}
\end{center}
\end{figure*}

\section{Low-entropy limit}\label{lowfilling}

The physics of this model is particularly transparent in the limit $m \to 1$, where either half-chain is almost fully magnetized. In this limit, most spins in the left (right) half-chain are in the $\uparrow$ ($\downarrow$) state, and a fraction $f \equiv \frac{1-m}{2}$ of the spins are in the minority state. When $f = 0$, one can regard the system as consisting of a single giant domain of $\uparrow$ spins above a vacuum of $\downarrow$ spins (or vice versa, by $\mathbb{Z}_2$ symmetry). The giant domain is immobile: $U(1)$ symmetry forbids it from growing or shrinking, while energy conservation only allows it to move on timescales exponential in the system size. For $f > 0$, one has some density $f^s$ of sequences (called ``strings'') of $s$ minority spins. In the integrable system we are considering, strings move with a characteristic velocity $v_s \sim {\rm e}^{-\eta (s-1)}$, where $\cosh \eta = \Delta$. 
(When $\Delta \gg 1$ this result follows from perturbation theory to order $s$.) When a string passes from one half-chain to the other, it does so by flipping its magnetization (e.g., going from a string of $\uparrow$ spins in a $\downarrow$ background to a string of $\downarrow$ spins in an $\uparrow$ background). By $U(1)$ symmetry a string of size $s$ must deposit magnetization $2s$ at the interface, thus shifting it by $2s$ in the direction opposite to the motion of the string (see Fig.~\ref{fig1}). 
The left and right half-chains are related by $\mathbb{Z}_2$ symmetry; thus, the two half-chains have identical distributions of strings~\cite{piroli2017}, and the domain undergoes unbiased Brownian motion due to its collisions with strings coming from either side. The diffusion constant $D_{\rm DW}$  associated with the motion of this domain wall can be computed from generalized hydrodynamics (GHD)~\cite{Doyon,Fagotti,JPA_GHDspecialissue} for any anisotropy $\Delta$ and filling $m$~\cite{dbd1,ghkv,dbd2}, using the results of Ref.~\cite{PhysRevLett.123.186601}. For $\Delta \to \infty$, it reads 
\beq
\underset{\Delta \to \infty}{\rm lim} D_{\rm DW} = \frac{4}{\pi \left(1 + 2 {\cosh} (2h) \right) },
\eeq
where we parametrized the filling as $m=\tanh h$. In the limit $m \to 1$, we have $D_{\rm DW} \sim \frac{4}{\pi} {\rm e}^{-2h}$, independently of the anistropy $\Delta$.

To relate the motion of the DW to $Q(t)$, we make the following key observation: \emph{whenever the DW moves away from the origin, $Q(t)$ goes up; whenever the DW moves toward the origin, $Q(t)$ goes down}. Thus the distribution of $Q(t)$ in this limit is simply the distribution of the \emph{distance} of the DW from the origin, i.e., the absolute value of a random walk:
\beq
P^{\rm OBC}_t(Q) = \frac{1}{\sqrt{\pi D_{\rm DW} t}} e^{-Q^2/(4 D_{\rm DW} t)}, \quad Q \geq 0.
\eeq
 The result above was obtained for a single interface, or open boundary conditions (OBC); for our setup involving PBCs we have two interfaces, and  $P_t$ is simply the convolution of $P^{\rm OBC}_t$ with itself since both interfaces are independent. We find
\beq\label{eqDistribution}
P_t({ Q}) = \sqrt{\frac{2}{\pi D_{\rm DW} t}} {\rm e}^{- \frac{{ Q}^2}{8 D_{\rm DW} t}}  {\rm erf} \left( \frac{ Q}{\sqrt{8 D_{\rm DW} t}}\right),
\eeq
where $ {\rm erf} x = \frac{2}{\sqrt{\pi}} \int_0^x {\rm e}^{-y^2} dy$ is the error function. 
It is straightforward to go from this expression to the generating function for FCS (at long times):
\beq\label{eqFCSXXZ}
\chi_t(\lambda) = -2 D_{\rm DW} t \lambda^2 + 2 \log \left[1+  {\rm erf} \left( i \sqrt{D_{\rm DW} t} \lambda \right) \right].
\eeq
The main features of our general result are apparent from this simple limit. {\bf First}, the expectation value $\langle {\hat Q}(t) \rangle \sim \sqrt{t}$. {\bf Second}, the width of the distribution \emph{also} grows as $\sqrt{t}$, so its variance grows as $t$. {\bf Third}, the distribution is strongly skewed, since it has a hard cutoff at the origin.
From the physical picture outlined above, the reason for this breakdown of the expected large deviation form~\eqref{eqLargedeviation} is straightforward: unlike ``standard'' diffusion, where there are many independently moving particles, in the present case the magnetization transfer is associated with a \emph{single} diffusing object, i.e., the interface. Thus the mean and variance are related as they would be for a \emph{single} random walker, not for an ensemble of random walkers. 

It is worth contrasting our result in this limit with what one would obtain, e.g., for a chaotic Hamiltonian or random unitary circuit with the same initial condition and a $U(1)$ conservation law~\cite{kvh,rpv}. (In the latter case the FCS is precisely that of the symmetric exclusion process, and its generating function is known exactly~\cite{Derrida:2009aa}.) In the chaotic case, each snapshot would reveal two polarized ``banks'' separated by a region of size $\sqrt{t}$ in which the domain wall has essentially melted and the state is random. By contrast, in the integrable case, each snapshot at time $t$ would reveal two highly polarized banks separated by a melted region of size $O(1)$, i.e., a sharp wall even at late times. That this domain wall stays sharp at late times is a consequence of integrability. Even though the domain wall remains sharp, its \emph{location} is random from shot to shot (as it depends on the initial positions of all the magnons in the sample). Thus, \emph{after averaging} the chaotic and integrable diffusive systems give the same spin profile, but the mechanisms are completely different: in chaotic case the domain wall melts in each branch of the wavefunction; in the integrable case, the domain wall remains sharp but wanders. These mechanisms naturally lead to very different statistics for the fluctuations. 

\section{General density} \label{general_filling}

We now turn to general filling $m$. The main difference in this case is that when the domain wall is at $x$, there is a finite density of strings in the interval $[0,x]$. The presence of these strings modifies the relationship between the domain wall position $x(t)$ (which still follows a random walk) and the magnetization transfer. Following the semi-classical picture outlined above, we have the microscopic relation $Q(t) = m_{[0,x]} x(t)$, where $m_{[0,x]}$ is the magnetization density in the interval between the domain wall and the origin. On average, this is just the equilibrium magnetization density on each side of the domain wall: it has an average value $m \times \mathrm{sgn}(x)$, as well as fluctuations given by standard thermal equilibrium statistical mechanics $ \langle \delta m_{[0,x]}^2  \rangle = \chi / |x|$, with $\chi = 1 - m^2$ the spin susceptibility in our infinite temperature ensemble. 

\subsection{Nonequilibrium case $m>0$: long time limit}

For any magnetization imbalance in the initial state $m>0$, we have $ m_{[0,x]}  \simeq m$ almost surely at long enough times, and the results of the previous section for $m \to 1$ apply immediately up to rescaling $Q \to Q/m$. A similar scaling with $m=\tanh h$ was found in the easy-plane regime $\Delta<1$ for the same domain wall initial state~\cite{PhysRevB.97.081111}, although transport is ballistic in this regime. 
We will show below that this factor reproduces all limiting cases, and captures properly the crossover to equilibrium ($m=0$). 

 In particular, we find the average magnetization transfer 
 \begin{equation}
 \langle \hat{Q} \rangle = 4 m \sqrt{\frac{D_{\rm DW} t}{\pi}},
 \end{equation}
 where $D_{\rm DW}$ can be computed from GHD for any $\Delta$ and $m$. The predicted dependence of $\langle \hat Q \rangle/\sqrt{t}$ on $m$ and $\Delta$ is plotted in Fig.~\ref{balazsfig}(d). This is consistent with the average magnetization obeying a diffusion equation 
 \begin{equation}
 \partial_t \langle S^z \rangle = D_{\rm DW} \partial_x^2 \langle S^z \rangle,
 \end{equation}
so we identify the spin and DW diffusion constants $D = D_{\rm DW}$, which is indeed known to be correct for the XXZ spin chain at half-filling $m\to 0$~\cite{dbd2,PhysRevLett.122.127202}, and is also believed to hold in general~\cite{JacopoPrivateComm}.
As $m \to 0$, this formula is consistent with linear response: for a single interface, we predict $\langle \hat{Q} \rangle \sim 2 m \sqrt{\frac{D_{\rm DW} t}{\pi}}$ with $m = \tanh h \sim h \ll 1$. In this regime, we have the linear response relation~\cite{PhysRevLett.122.210602,2107.00038} for a small domain wall: $\langle S_z(x,t) \rangle = \frac{h}{2} - 4 h \int_{-\infty}^x dy \langle S_z(x,t)  S_z(0,0) \rangle$, where $ \langle S_z(x,t)  S_z(0,0) \rangle = \frac{1}{8 \sqrt{\pi D t}} {\rm e}^{- x^2/(4 D t)} $ is evaluated in the equilibrium infinite temperature state, and $D$ is the spin diffusion constant at half-filling~\cite{dbd2,PhysRevLett.122.127202,gvw}. Using this linear response relation and Eq.~\eqref{eqP}, we find $\langle \hat{Q} \rangle/ h \sim 2 \sqrt{\frac{D t}{\pi}}$ which agrees with our prediction.
Higher cumulants follow from our main result~\eqref{eqDistribution} up to rescaling by $m$, with the variance $\langle {\hat Q}(t)^2 \rangle_c \sim t$ and a universal skewness $ \frac{4 -\pi}{(\pi - 2)^{3/2} } \simeq 0.704\dots$  

\subsection{Equilibrium limit and finite time crossovers}

We now discuss the crossover to the equilibrium state at $m = 0$. Here, by symmetry, there can be no net magnetization transfer: this immediately follows from the discussion above. Furthermore, the variance of the magnetization transfer scales as $\sqrt{t}$ with a prefactor that is independent of $m$. One can see this from our discussion above:  over a time $t$, the domain wall moves over a typical distance $x \sim \sqrt{t}$, but the magnetization density $m_{[0,x]}$ the interval between the domain wall and the origin is 0 on average, with typical fluctuations $\sim \frac{1}{\sqrt{x}}$. Therefore,  the net magnetization transported only scales as $\sqrt{x} \sim t^{1/4}$ as one would expect from central limiting behavior. For small but nonzero $m$, the variance crosses over from an $m$-independent piece that scales as $t^{1/2}$ to an $m$-dependent piece that scales as $m^2 t$. The crossover timescale thus scales as $t_\star \sim 1/m^4$ and rapidly diverges near $m = 0$.

\begin{figure}[tb]
\begin{center}
\includegraphics[width = 0.45\textwidth]{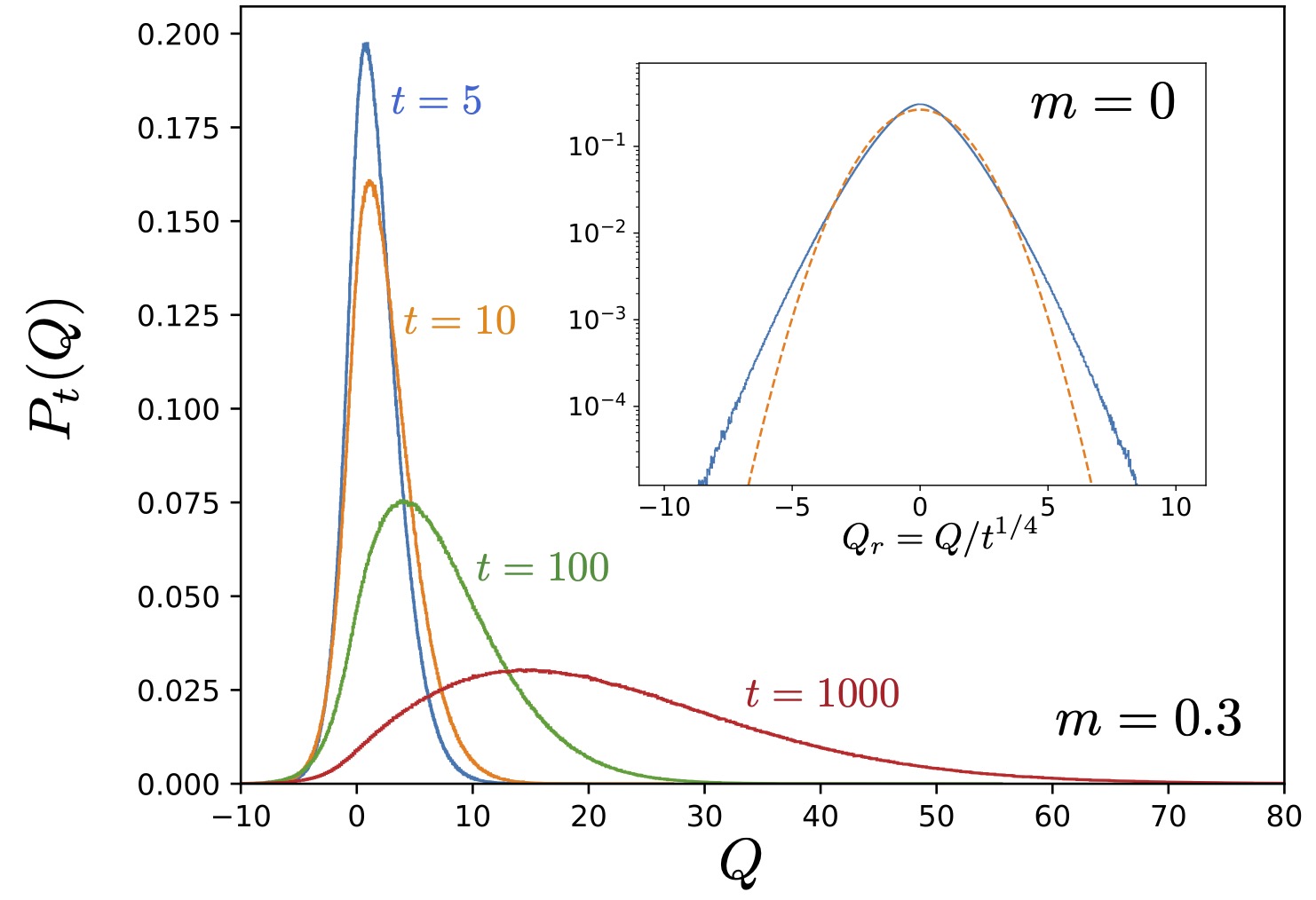}
\caption{Finite time crossover of the spin transfer distribution for $m>0$, obtained from the PBC analog of eq.~\eqref{eqFullDOBC}. Here we set $D=1$, and the distribution was obtained from sampling the integrals over $x$ and $m_{[0,x]}$. Inset: equilibrium distribution ($m=0$) of the rescaled transfer $Q_r = Q /t^{1/4}$ for PBC. The orange dashed line is a Gaussian distribution with the same variance for comparison.}
\label{figCrossover}
\end{center}
\end{figure}

\begin{figure}[tb]
\begin{center}
\includegraphics[width = 0.45\textwidth]{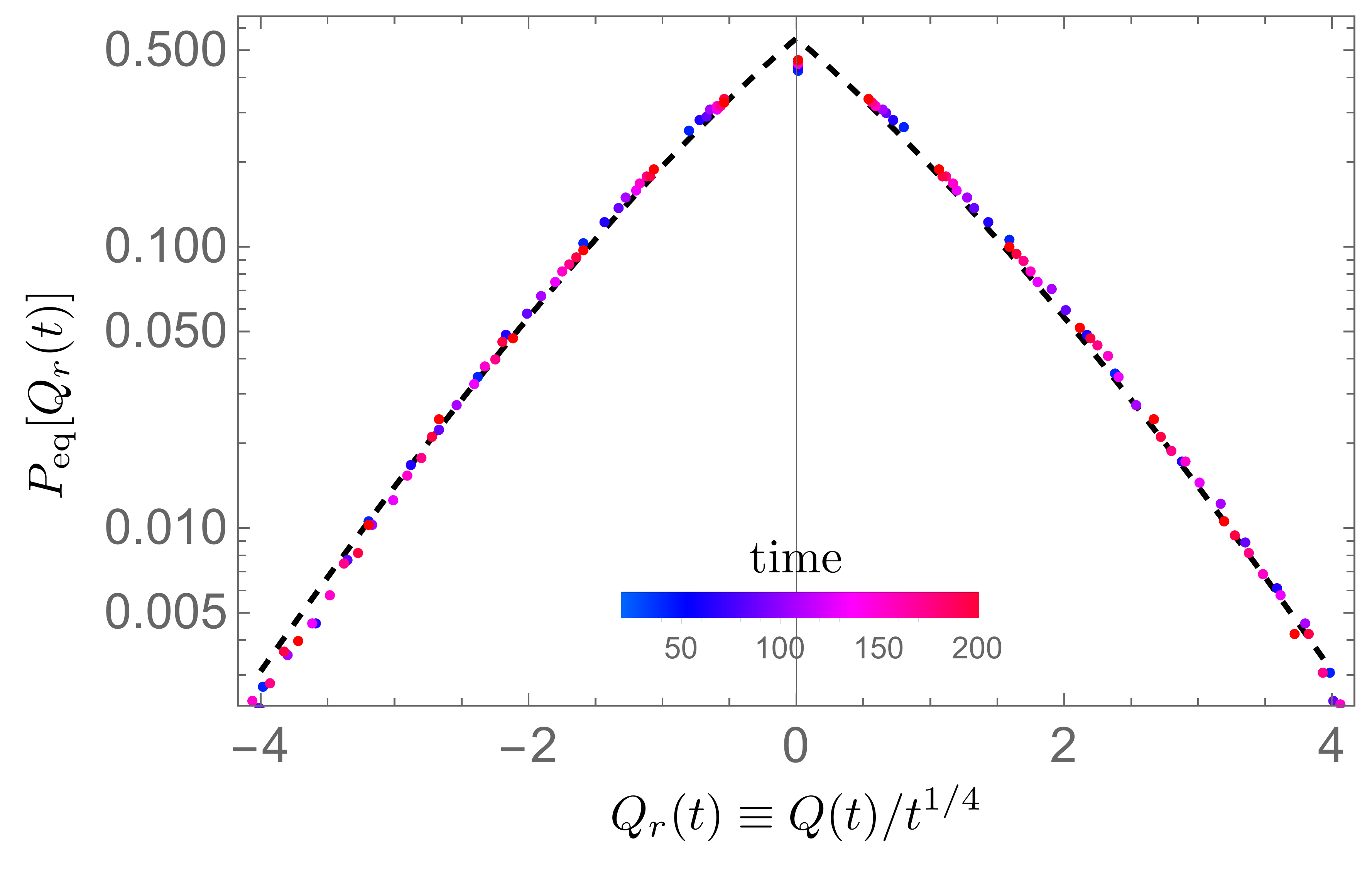}
\caption{Rescaled distribution of magnetization transfer for \emph{open boundary conditions} (OBC) in equilibrium (no bias, $m=0$). The dashed line is the theoretical prediction~\eqref{eqFullDOBC} with $m=0$, and the dots are rescaled histograms at various times from $t = 20$ to $t = 200$. The histograms are over $60000$ realizations.}
\label{figeqdistribution}
\end{center}
\end{figure}

Our picture above captures this crossover quantitatively, as well as the equilibrium distribution. Recall that for a single interface (OBC), we have $Q(t) = m_{[0,x]} |x(t)|$ with $x(t)$ normally distributed with variance $2 D_{\rm DW} t$ and zero mean, and $m_{[0,x]}$ a normally distributed variable with mean $m$ and variance $(1-m^2)/|x|$. 
Writing $P^{\rm OBC}_t(Q)$ in terms of the random variables $m_{[0,x]}$ and $x$ with a delta function $\delta(Q-m_{[0,x]} |x|(t))$ enforcing the constraint gives:
\begin{equation} \label{eqFullDOBC}
P^{\rm OBC}_t(Q)=\int_0^\infty dx \frac{\exp \left(-\frac{x^2}{4 D_{\rm DW} t}-\frac{x \left(\frac{Q}{x}-m\right)^2}{2 \left(1-m^2\right)}\right)}{\sqrt{2 \pi^2 (1-m^2)} \sqrt{D_{\rm DW} t x}}.
\end{equation}
Using this expression, we can easily compute the FCS~\eqref{eqFCS} in closed form. Going back to PBC (corresponding to multiplying the FCS by a factor of two since there are two independent interfaces), we find
\begin{align}\label{eqFCSXXZfull}
\chi_t(\lambda) &= -2 D_{\rm DW} t \tilde{\lambda}^2 + 2 \log \left[1+  {\rm erf} \left( i \sqrt{D_{\rm DW} t} \tilde{\lambda} \right) \right], \notag \\
\tilde{\lambda} &= m \lambda + i \frac{(1-m^2)\lambda^2}{2}.
\end{align}
Clearly in the limit $m \to 1$, we recover~\eqref{eqFCSXXZ}. The corresponding probability distribution $P_t(Q)$ is plotted in Fig.~\ref{figCrossover}.
For any $m>0$, the distribution of magnetization transfer broadens symmetrically at short times. This broadening is a subleading effect and the late-time distribution still approaches the universal form~\eqref{eqFCSXXZ} (with $\lambda \to m \lambda$) that we discussed in the $m\to 1$ limit. This can be seen by letting $t \to \infty$ and $\lambda \to 0$ with $ t \lambda^2$ fixed, so that in that limit, $t \tilde{\lambda}^2 \to t \lambda^2 $. The formula~\eqref{eqFCSXXZfull} can be used to fully characterize the finite-time crossover for finite $m$. For example, the variance reads
\begin{equation}
 \langle \hat{Q}^2 \rangle_c=  4 \left(1-m^2\right) \sqrt{ \frac{D_{\rm DW} t}{\pi }}+ 4 D_{\rm DW} m^2 (1 -2/\pi) t .
\end{equation}
As we anticipated above,  the variance crosses over from $\sqrt{t}$ to $m^2 t$ on a time scale $t_\star \sim 1/m^4$. In the equilibrium case $m=0$, we predict 
\begin{equation}
\chi^{\rm eq}_t(\lambda)=\frac{1}{2} D_{\rm DW} \lambda ^4 t +  \log \left[1-\text{erf}\left(\frac{\lambda ^2 \sqrt{D_{\rm DW} t} }{2} \right)\right]^2,
\end{equation}
with the corresponding probability distribution plotted in the inset of Fig.~\ref{figCrossover}.
A closely related equilibrium distribution was discussed in the context of an exactly solvable cellular automaton in Ref.~\cite{krajnik2022exact}.

\section{Numerical support} \label{numerics}

In general the quantum XXZ spin chain must be simulated using numerical methods that are restricted to modest times and system sizes. To compare our predictions to numerics that we can converge to very late times, we have performed simulations on an integrable cellular automaton (the ``folded XXZ automaton" from Sec.~\ref{setup}) that mimics the behavior of the XXZ model in the large-$\Delta$ limit~\cite{balazs1, balazs2}. The automaton dynamics is made up of four-site updates such that the states on the middle two sites are swapped conditional on the first and last site being in the same state. This rule conserves the total number of domain walls and of $\uparrow$-spins; thus it is evidently closely related to the $\Delta \to \infty$ limit of the XXZ model~\cite{10.21468/SciPostPhysCore.4.2.010}. If these updates are applied in the pattern described in Ref.~\cite{balazs1} this model can be shown to be integrable. As the upper panel of Fig.~\ref{balazsfig} shows, $P_t({Q})$ in the automaton convincingly approaches the prediction~\eqref{eqDistribution} at late times, even starting from states that are relatively far from unit filling. Moreover, the mean and standard deviation of the magnetization transfer clearly both scale as $\sqrt{t}$. In equilibrium, the rescaled distribution of magnetization transfer matches our prediction~\eqref{eqFullDOBC} with $m=0$ (Fig.~\ref{figeqdistribution}). 

We have also checked our predictions against a fully quantum XXZ model, whose dynamics we simulate exactly. The initial state is a direct product of random pure states with definite total magnetization on each half-system so that the magnetization transfer is well defined even at early times. To make the simulations more tractable we use an integrable Trotterization~\cite{PhysRevLett.121.030606, Ljubotina-Prosen2019_ballistic} of the Hamiltonian model made up of two-spin gates $U=\exp [i (S^x S^x + S^y S^y + \Delta S^z S^z)]$ arranged in a brickwork pattern with periodic boundaries, and run our simulations on a GPU using JAX~\cite{Morningstar-Vidal2022tpu, JAX}. We set $\Delta = 5$, and measure time in units of the period of the quantum circuit (two layers of gates lasts one unit of time). At each time we extract the distribution $P_t(Q)$ and average it over 50 samples. {Unfortunately, the combination of initial transients and finite-size/time effects make it so that we cannot conclusively establish how the variance of $Q$ scales with time, or what the skewness does at late times.} Nonetheless, the observed behavior is consistent with our theoretical predictions [Fig.~\ref{balazsfig}(e)], and the skewness clearly follows a growth similar to what occurs in the automaton [Fig.~\ref{balazsfig}(f)].

\begin{figure*}[tb]
\begin{center}
\includegraphics[width=.95\textwidth]{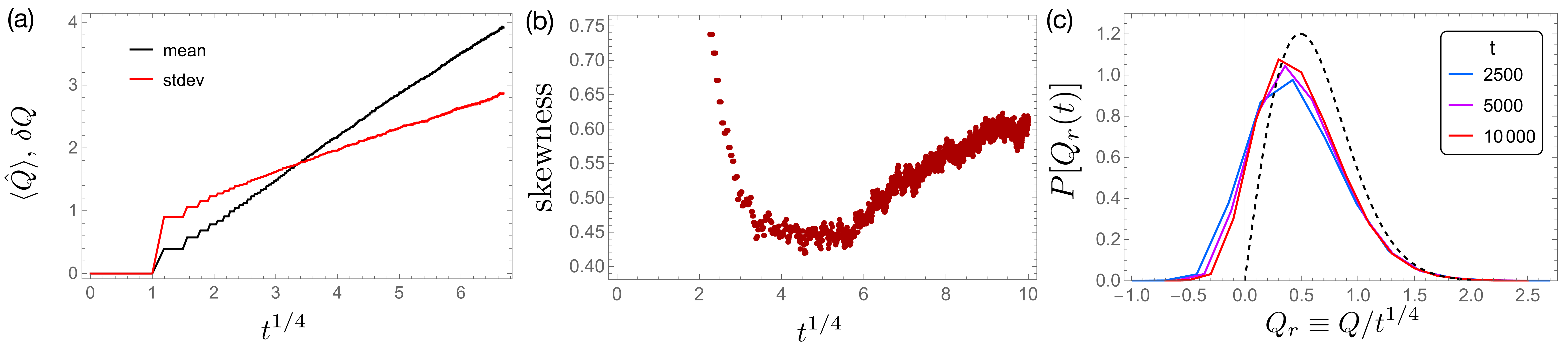}
\caption{Data on the stochastic XNOR model. (a, b)~Mean, standard deviation, and skewness of magnetization transfer, vs. $t^{1/4}$. (c)~Rescaled distribution of magnetization transfer, compared to the infinite-time prediction (dashed line). All simulations involved initial states at $f = 5/6$, averaged over $40000$ realizations, each with a different initial state and noise pattern.}
\label{xnorfig}
\end{center}
\end{figure*}

We now briefly present results on the XNOR model, discussed in more detail in Ref.~\cite{PhysRevLett.127.230602,2021arXiv210913251D}. The dynamics of this model consists of four-site updates in which the middle two spins can flip flop if the two outer spins are aligned. When these updates are applied randomly in space and time, one ends up with a stochastic model in which magnons move diffusively but the number of magnons and frozen domains is separately conserved~\cite{PhysRevLett.124.207602,PhysRevLett.125.245303}. Integrability breaking in the large-$\Delta$ limit gives rise to XNOR dynamics over a parametrically large time window controlled by the parameter $\Delta$. 

The discussion in previous sections also applies to this model with some minor changes. First, the domain wall moves \emph{subdiffusively}. Over a time $t$ the domain wall collides with magnons coming from either side that began within a distance $\sim \sqrt{t}$ of the domain wall. The imbalance between left and right movers is $t^{1/4}$, and correspondingly that is also the scale over which the domain wall moves. Since the motion of the domain wall is what controls the magnetization transfer in this model, the mean and standard deviation are expected once again to scale the same way, and the distribution is expected to develop a large skewness. In fact, there is strong numerical evidence~\cite{PhysRevLett.127.230602} that the domain wall moves with a rescaled Gaussian, so the entire generating function for the full counting statistics in this model is expected to be identical to that discussed in earlier sections, except for the change in the dynamic scaling.

In Fig.~\ref{xnorfig} we test these arguments against numerical evidence. The data quality for this model is worse than for the integrable model discussed in the main text, primarily because the much slower dynamics makes it hard to go out to late enough times to see the asymptotic physics. Nevertheless, the first few moments behave very similarly to those for the integrable model. A look at the full distribution $P_t(Q)$ (Fig.~\ref{xnorfig}(c)) indicates the reason for the slow convergence. In a state at finite filling the ``domain wall'' is often some distance behind the origin (e.g., because there is a large density of quasiparticles between the domain wall and the origin). Eventually, as the domain wall spreads out, these frozen-in initial fluctuations become irrelevant, as they have some time-independent width while the rest of the distribution broadens with time. Thus the part of the distribution left of the origin becomes negligible, and the distribution approaches the theoretical prediction, although this approach is slow [Fig.~\ref{xnorfig}(d)].

\section{Discussion} \label{conc}

We have presented an explicit form for the distribution function of the magnetization transfer, $P_t(Q)$, for the domain wall initial state in the XXZ model. We showed that even though magnetization transfer is diffusive on average, the distribution corresponds to a distinct non-Gaussian universality class. 
We have argued, and presented numerical evidence, that this distribution is universal at asymptotically late times for \emph{any} nonzero imbalance; our arguments also imply that it should hold asymptotically for any nonzero easy-axis anisotropy. While we used results from integrability, in fact the main phenomenological features survive even in the absence of integrability, at large $\Delta$, because they only rely on the stability of magnons. When integrability is broken at large $\Delta$, there is a parametrically large time window (polynomial in $1/\Delta$~\cite{2021arXiv210913251D}) for which magnons are diffusive but stable; in this time window, one expects \emph{subdiffusion}~\cite{PhysRevLett.127.230602,2021arXiv210913251D} with anomalous counting statistics. In this regime, the universal skewed distribution~\eqref{eqDistribution} once again arises asymptotically; however, its width is now set by $t^{1/4}$ rather than $t^{1/2}$, as discussed above. At finite $\Delta$, this anomalous counting statistics arises at intermediate times, and presumably eventually crosses over to regular diffusion. 

Our work shows that systems with the same hydrodynamics can nevertheless belong to distinct universality classes, and these distinctions can be exposed using observables that are natural to near-term simulators. 
Our predictions can be tested in various currently feasible experiments. Single-site-resolved microscopy has been demonstrated and extensively used to study dynamics, including in the isotropic Heisenberg model~\cite{2107.00038}. The anisotropic XXZ spin chain was experimentally realized and studied in Ref.~\cite{Jepsen:2020aa}; while this setup did not involve site-resolved microscopy, such an extension is experimentally feasible. Site-resolved imaging capabilities are also built into realizations of programmable quantum matter including Rydberg atom arrays~\cite{bernien2017probing} and superconducting qubit arrays~\cite{arute2019quantum}. The easy-axis XXZ model (or its Trotterized version) is straightforward to implement on these platforms. However, the extent to which our predictions can be checked depends on how well depolarizing noise can be mitigated. Depolarizing noise will contaminate the full counting statistics as it will lead to particle number fluctuations that are unrelated to transport. 

A particularly interesting question for future work is to extend our theoretical picture to the Heisenberg model, $\Delta =1$. A naive extension of our analysis would suggest that the magnetization transfer remains diffusive at the isotropic point, consistent with the numerics in Ref.~\cite{PhysRevLett.128.090604}. This is because $D_{\mathrm{DW}}$ remains finite at $\Delta = 1$ for $m \neq 0$~\cite{gvw}. The crossover to diffusion is predicted~\cite{gvw} to happen at times $t(m) \sim 1/m^3$, and thus diverges rapidly near zero bias. However, a subtlety that arises in the Heisenberg limit is that the dynamics does not freeze out even at $m = 1$. Therefore, to capture the Heisenberg case, one would have to augment our analysis of the ``Brownian'' contribution to magnetization transfer, coming from magnon--domain wall collisions, with a more careful analysis of the $m = 1$ dynamics. A related point is that the experiment in Ref.~\cite{2107.00038} measured a finite skewness for $P_t(Q)$ whose value was consistent with KPZ predictions, even though the experiment was performed with a nearly polarized initial domain wall, $m \approx 1$, which is outside the near-equilibrium regime ($m\rightarrow 0$) where KPZ physics is expected to hold. Our analysis in this work shows how $P_t(Q)$ can display finite skewness even in diffusive systems in the strongly anisotropic regime, and for nearly polarized initial conditions, far from any KPZ regime. An interesting question for future work is to determine whether the measured skewness at $\Delta = 1, m \rightarrow 1$ is indeed related to KPZ physics, or whether it has a distinct dynamical origin related to the considerations in this work.  We leave these questions to future work.

\acknowledgements{S.G. thanks I. Bloch, J. Kemp, M. Knap, F. Machado, K. Takeuchi, D. Wei, N. Yao, B. Ye, and J. Zeiher for collaborations on related work. 
{A.M. thanks Guifre Vidal for discussions of related work.} R.V. and S.G. thank J. De Nardis, E. Ilievski, Z. Krajnik and T. Prosen for comments on the manuscript, and for discussions on related works.  
This work was supported by the US Department of Energy, Office of Science, Basic Energy Sciences, under Early Career Award Nos. DE-SC0019168 (R.V.) and DE-SC0021111 (V.K), the Alfred P. Sloan Foundation through a Sloan Research Fellowship (R.V. and V.K.), the Packard Foundation through a Packard Fellowship in Science and Engineering (V.K.), the NSF through grant DMR-1653271 (S.G.), and the DARPA DRINQS progam (A.M.). Some simulations presented in this work were performed on computational resources managed and supported by Princeton Research Computing.
V.K. and S.G. thank the Kavli Institute of Theoretical Physics (KITP) for hospitality. KITP is supported in part by the National Science Foundation under Grant No. NSF PHY-1748958.

\bibliography{main}

\appendix

\section{Extended data from the XXZ model simulations}
For completeness, in this section we provide an extended dataset from simulations of the integrable XXZ circuit, with both periodic and open boundary conditions. In part of Fig.~\ref{fig:supp_evolution} we show an example of the evolution of the spatial magnetization profile, and distribution of magnetization transfer, over time, for a particular initial polarization $m=1/3$. In the same figure we also show the distribution $P_t(Q)$ at a particular time ($t=20$) for several initial polarizations and boundary conditions. Fig.~\ref{fig:supp_Q_stats} shows the mean, standard deviation, and skewness of $P_t(Q)$ for periodic and open boundaries and several initial polarizations $m$. As was stated in the main text, working with such small systems sizes for the fully quantum model limits our ability to conclusively extract late-time behavior, so the main purpose of this data is to check for consistency with our theoretical arguments and cellular automaton numerics.

\begin{figure*}
    \centering
    \includegraphics[width=0.32\textwidth]{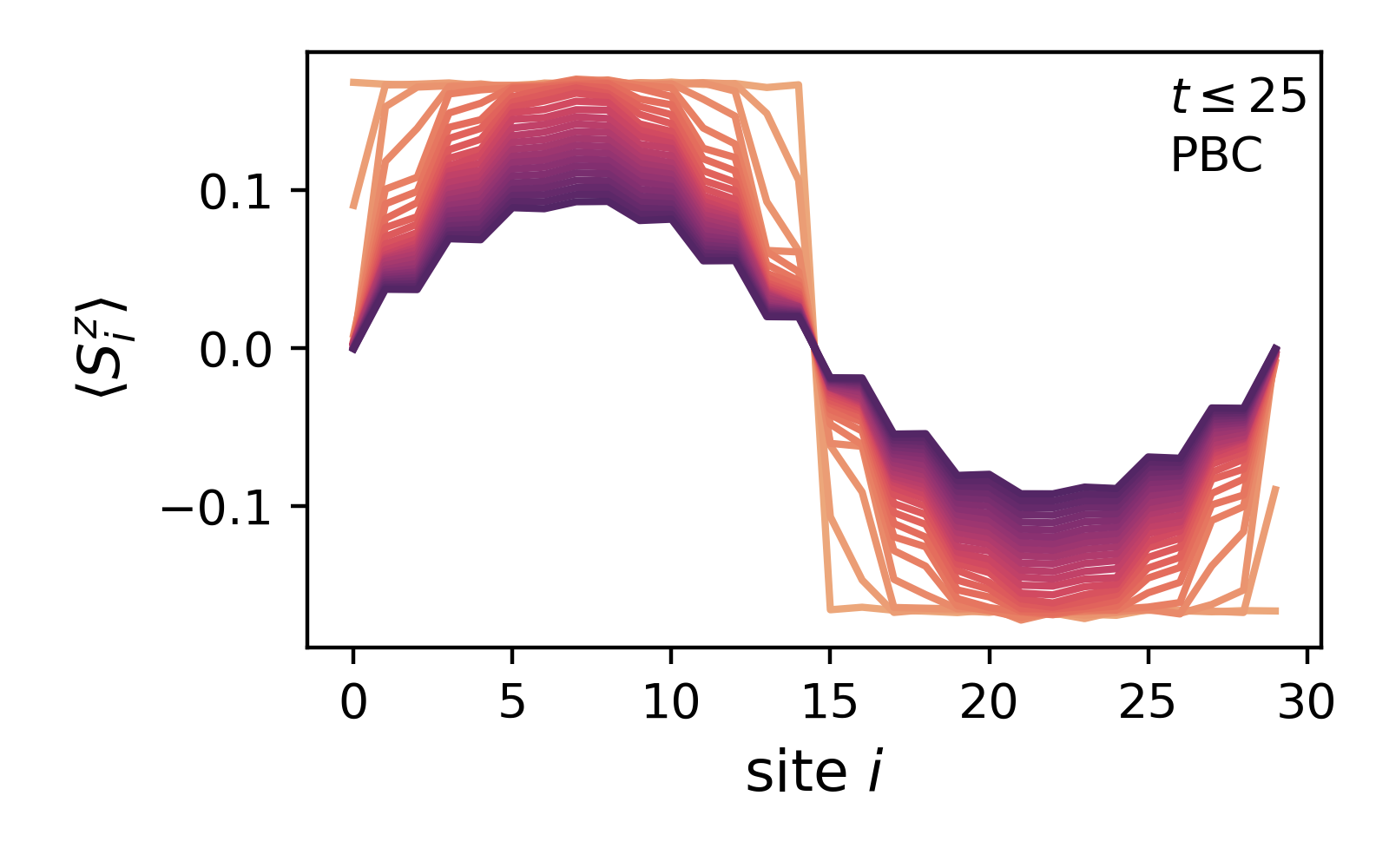}
    \includegraphics[width=0.32\textwidth]{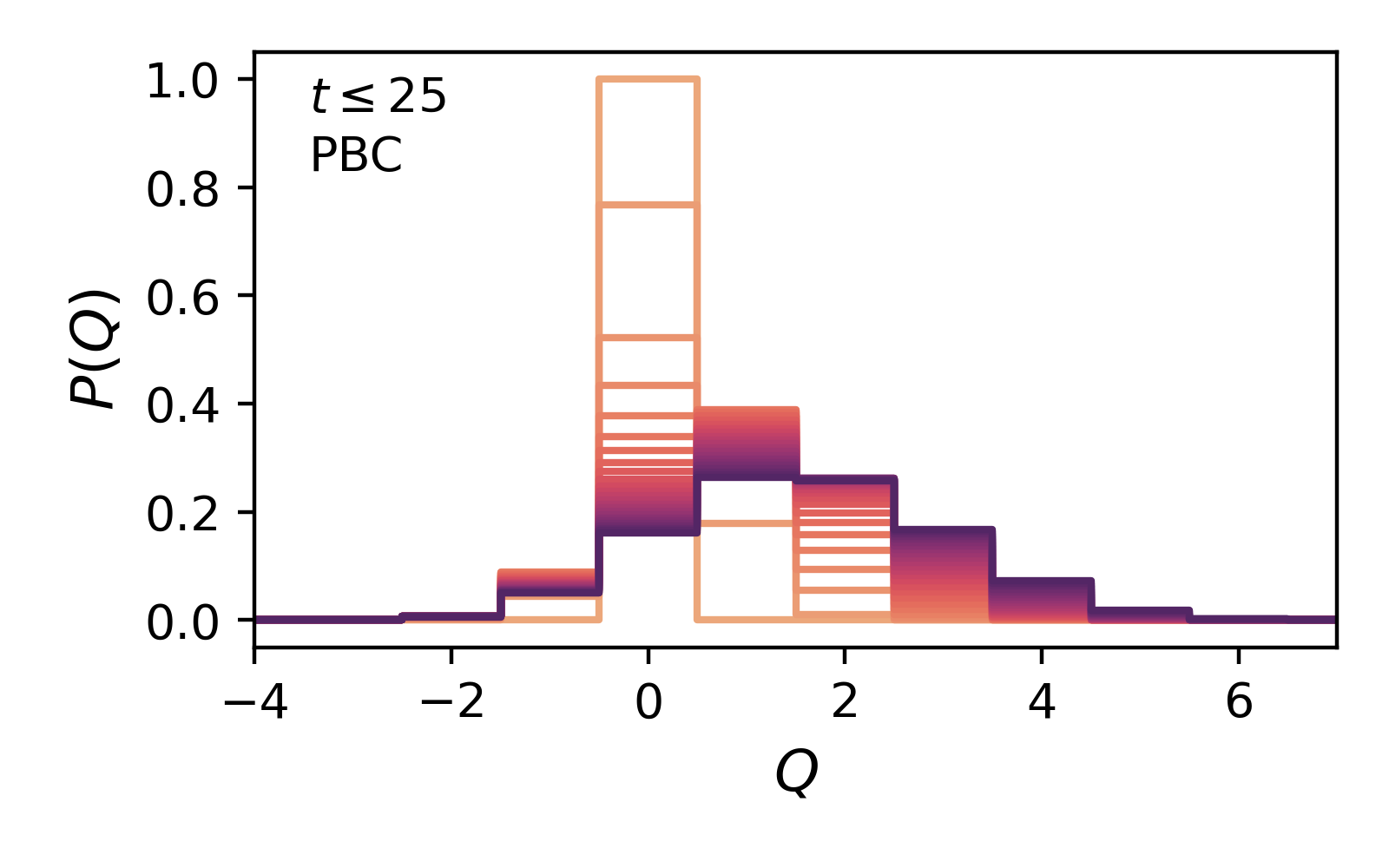}
    \includegraphics[width=0.32\textwidth]{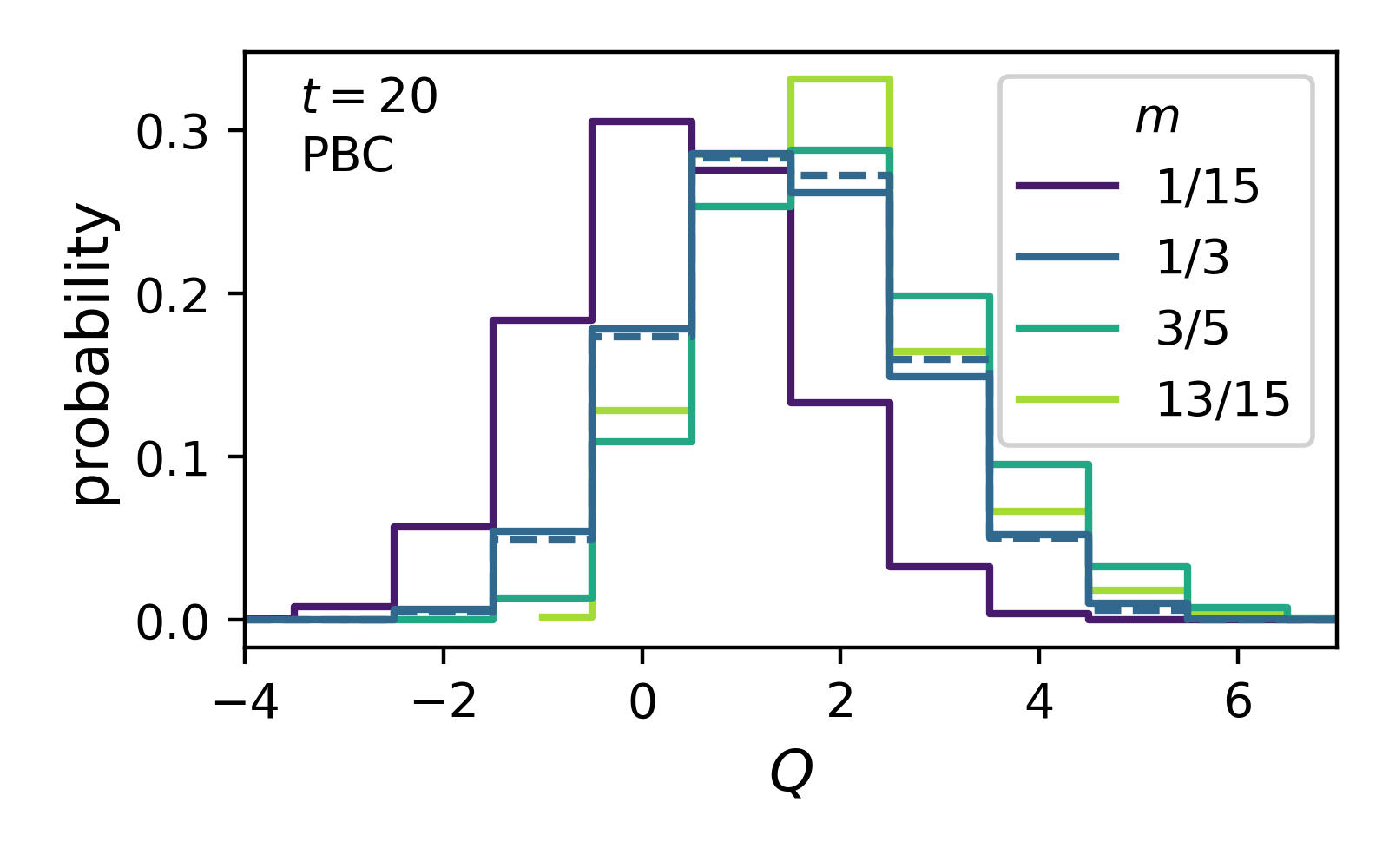}\\
    \includegraphics[width=0.32\textwidth]{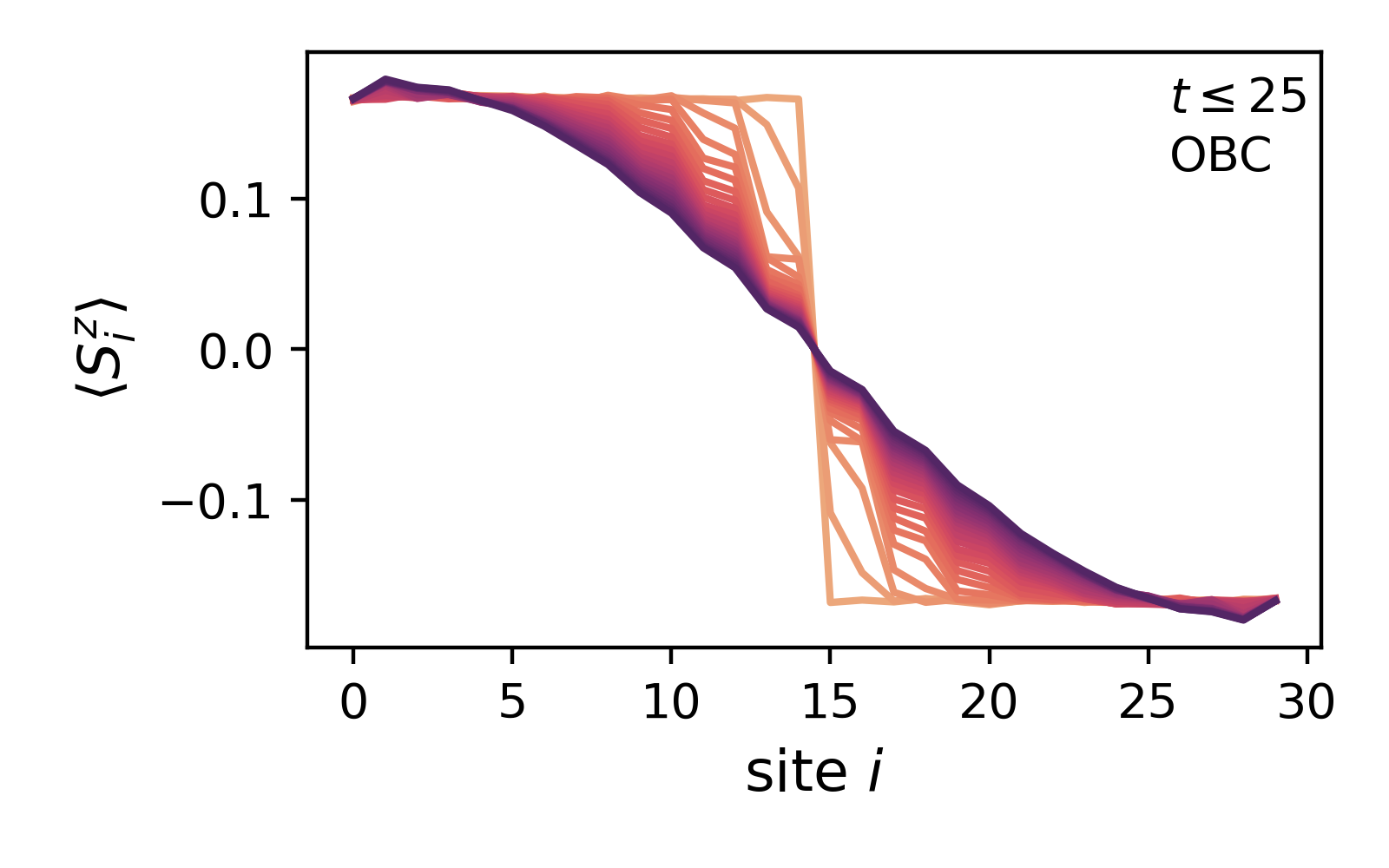}
    \includegraphics[width=0.32\textwidth]{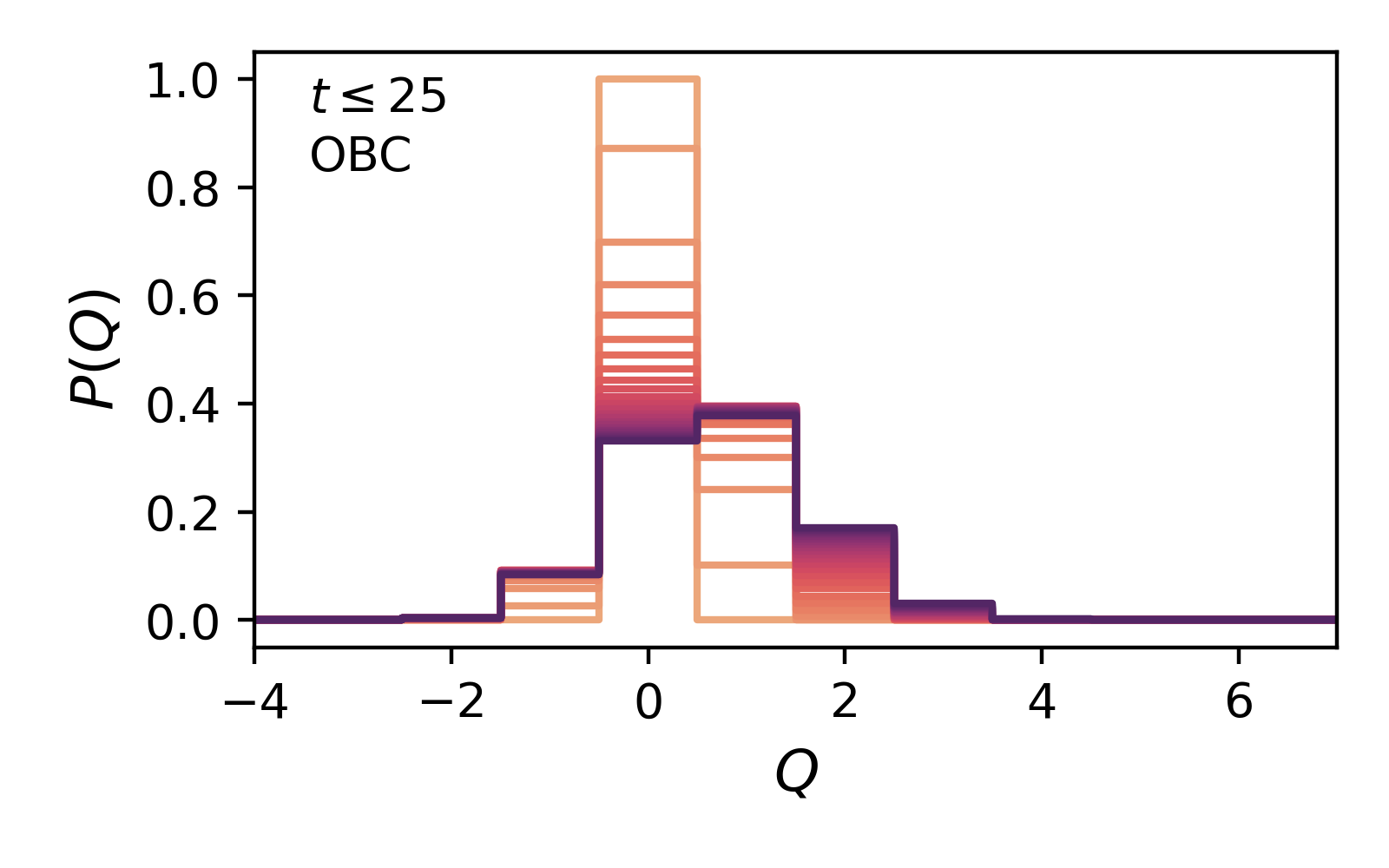}
    \includegraphics[width=0.32\textwidth]{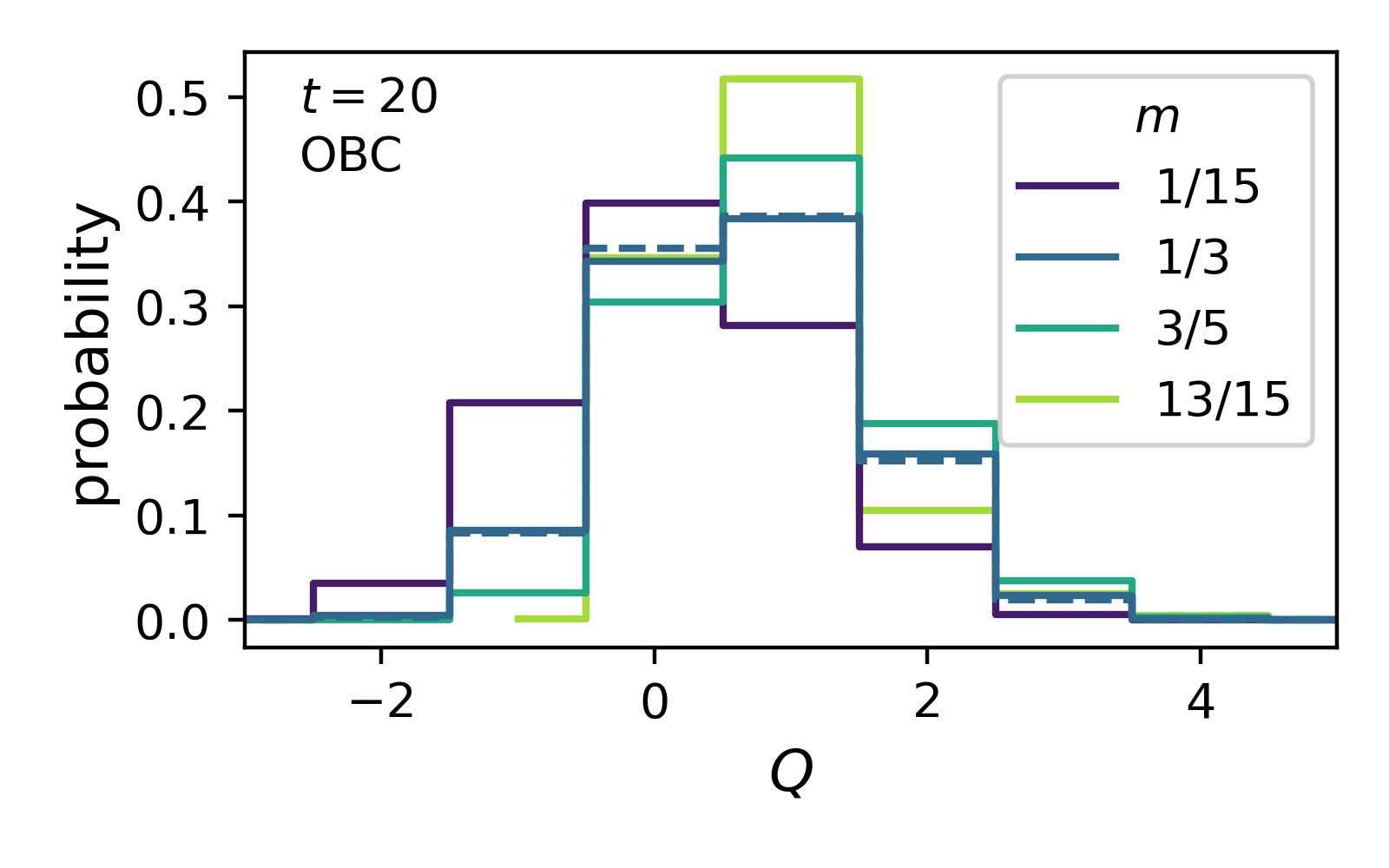}
    \caption{The spatial magnetization profile $\langle S^z_i \rangle$, and distribution of magnetization transfer. The top (bottom) panels are for periodic (open) boundaries. Time $t\in [0, 25]$ is represented by light to dark curves in the first two columns, which are for $N=30$ spins and $m=1/3$. In the final column we show the quantum distribution over $Q$ at time $t=20$, solid lines represent simulations with $N=30$ spins, and the dashed line is at $N=24$ and $m=1/3$.}
    \label{fig:supp_evolution}
\end{figure*}

\begin{figure*}
    \centering
    \includegraphics[width=0.32\textwidth]{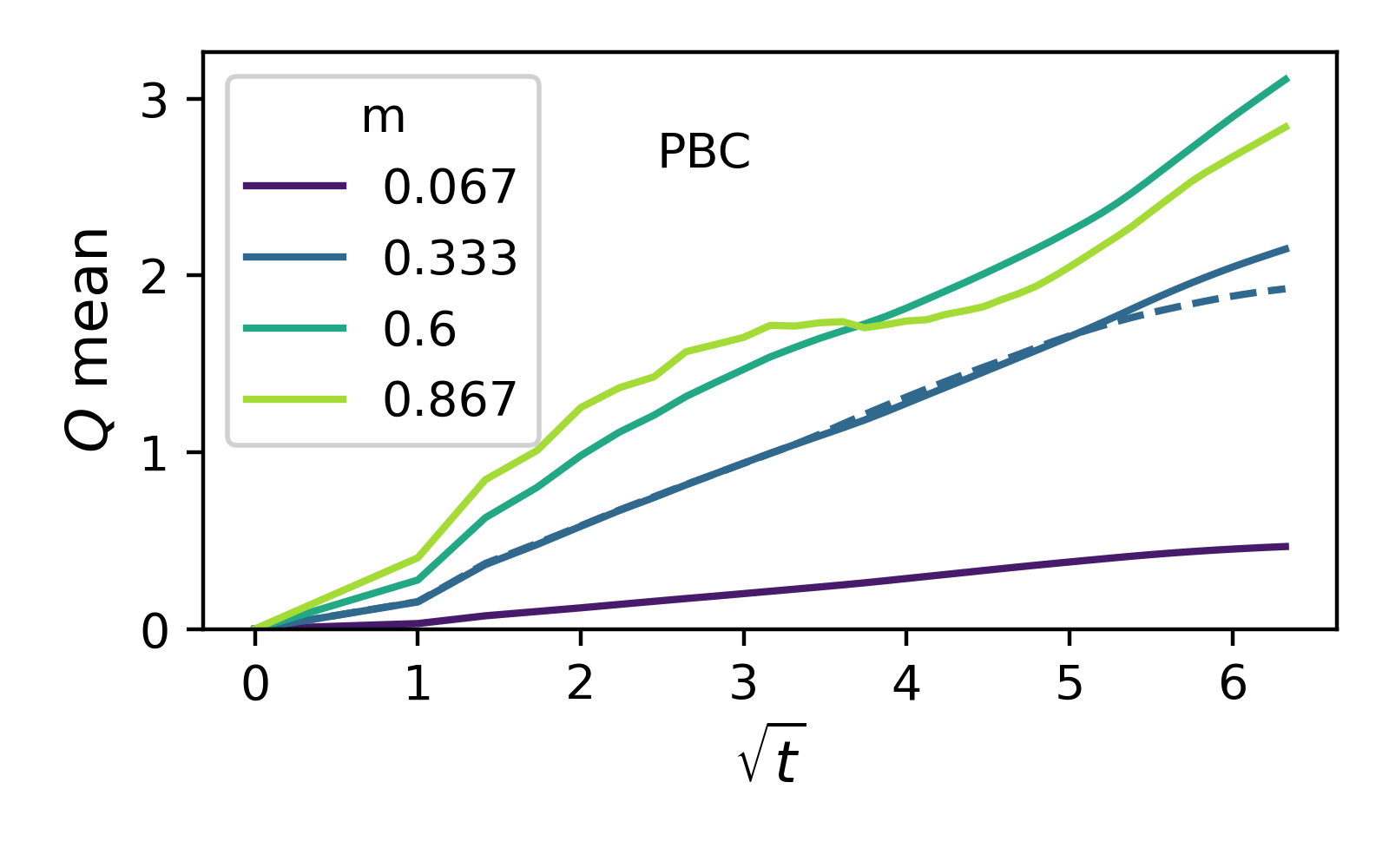}
    \includegraphics[width=0.32\textwidth]{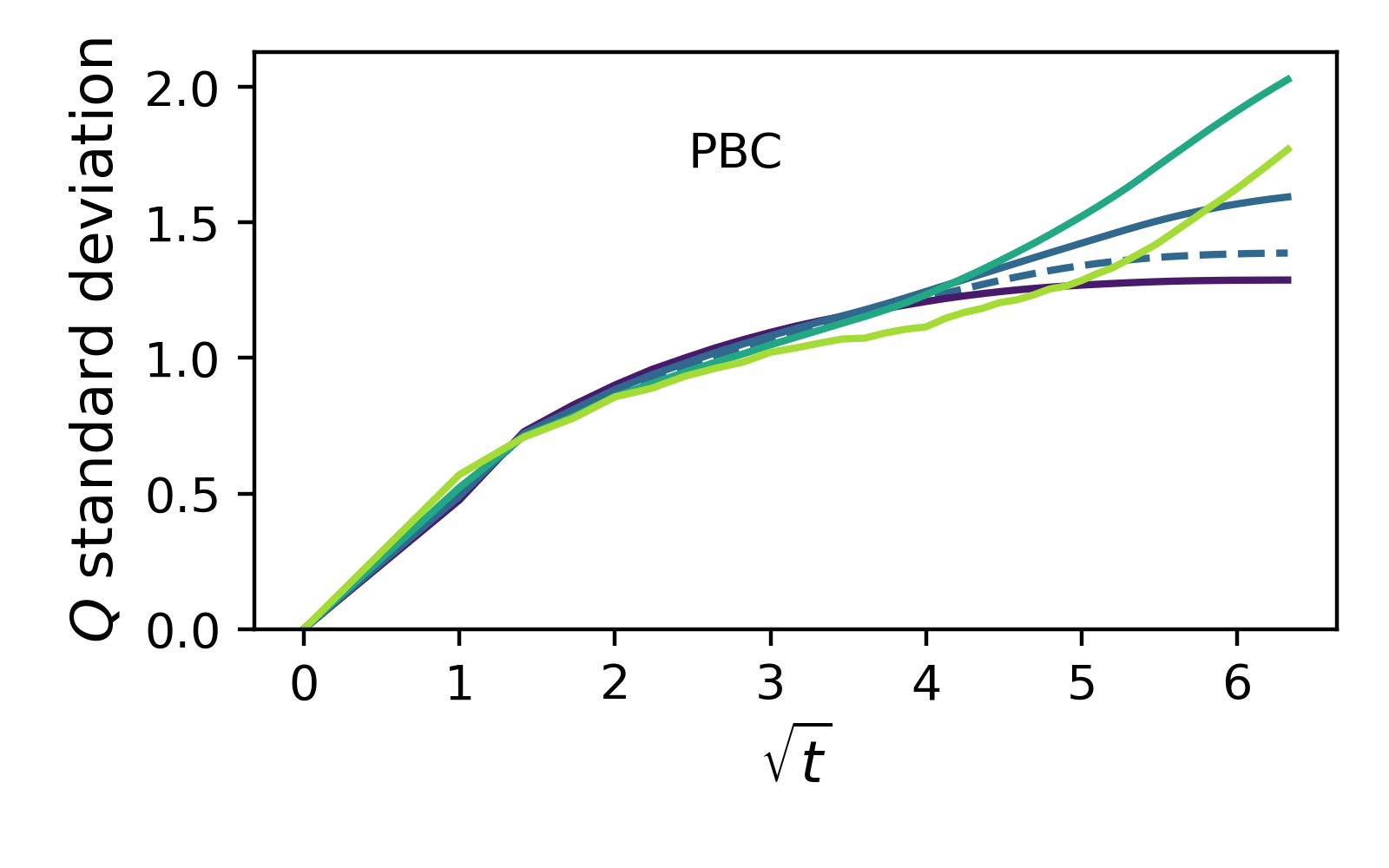}
    \includegraphics[width=0.32\textwidth]{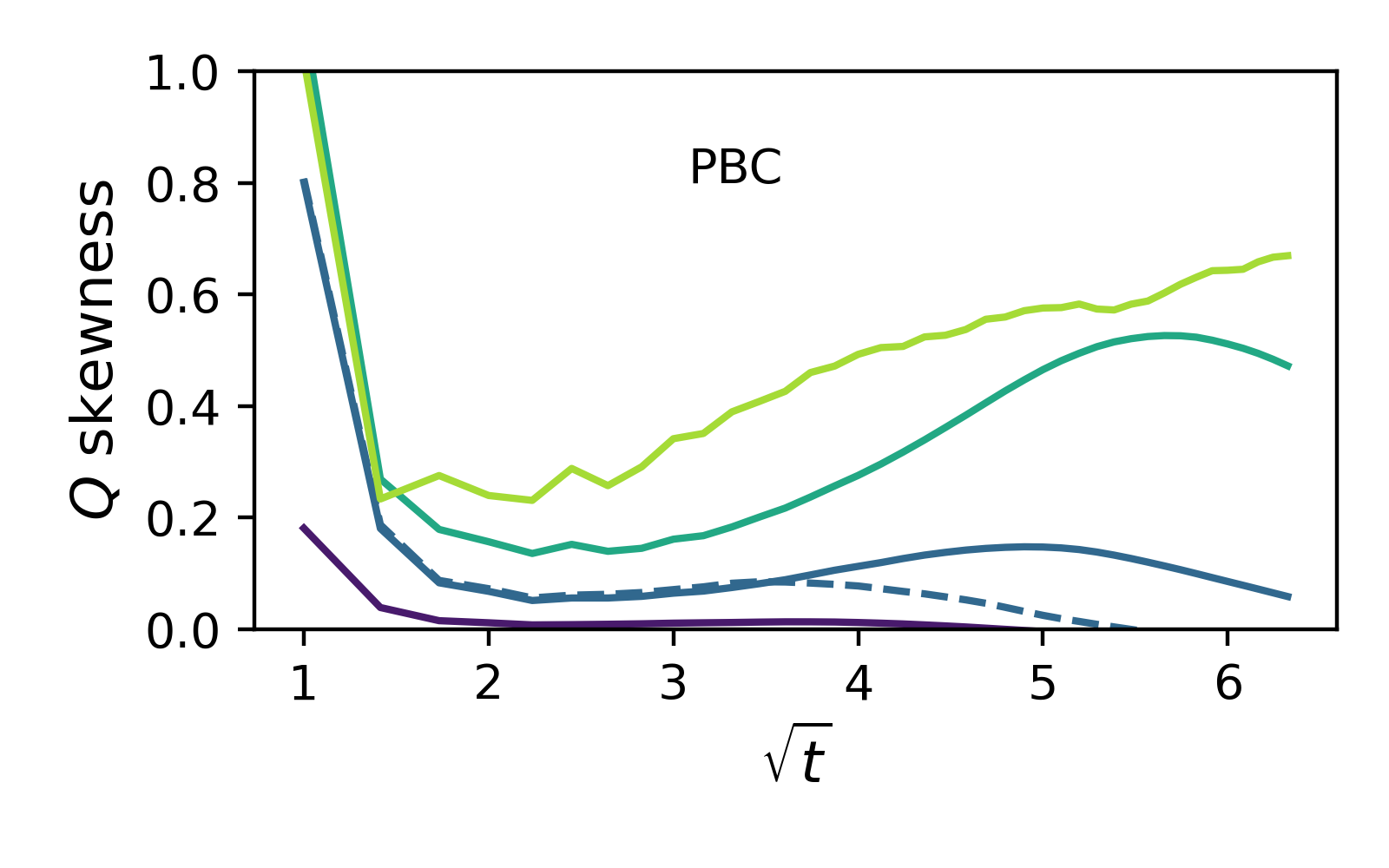}\\
    \includegraphics[width=0.32\textwidth]{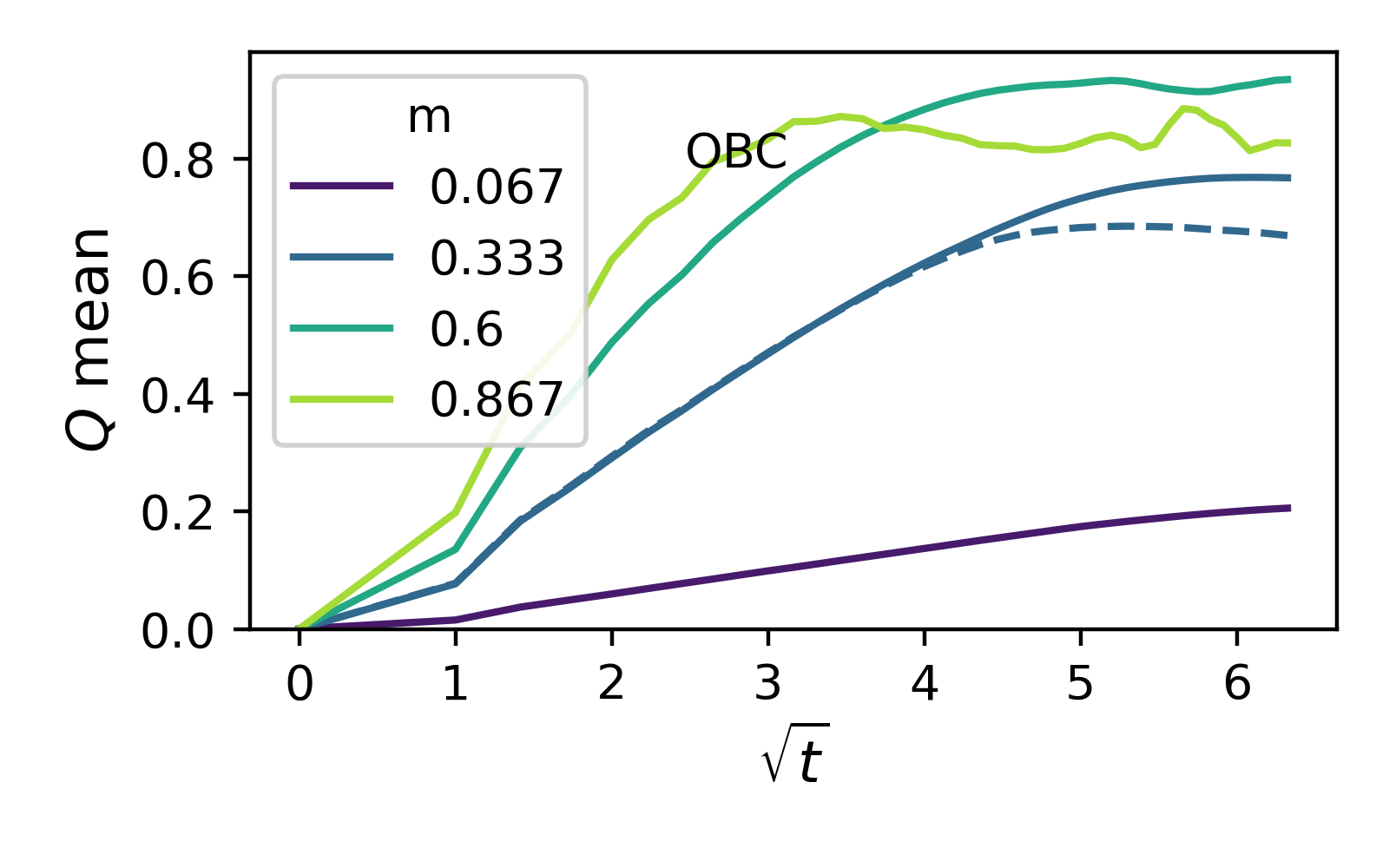}
    \includegraphics[width=0.32\textwidth]{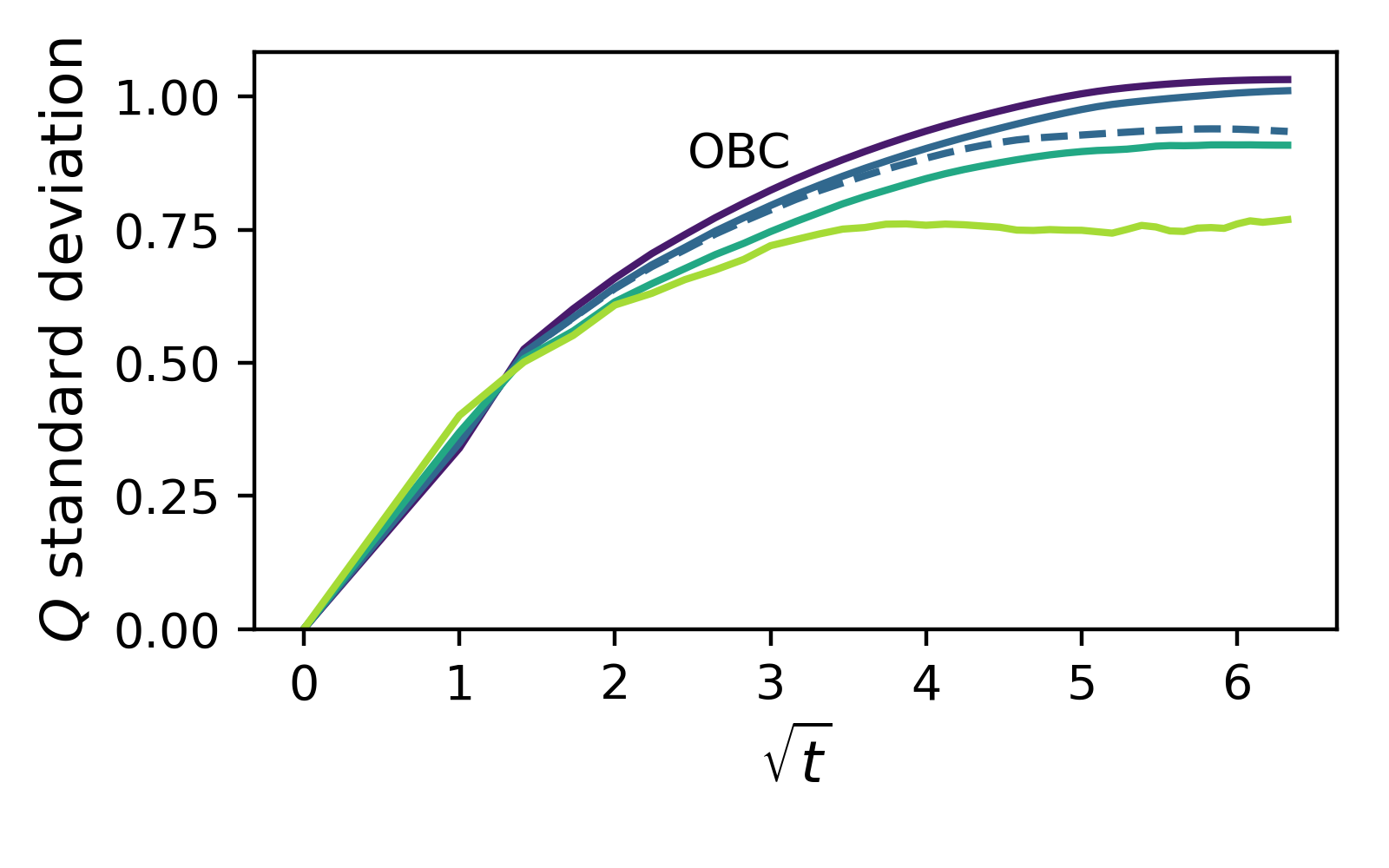}
    \includegraphics[width=0.32\textwidth]{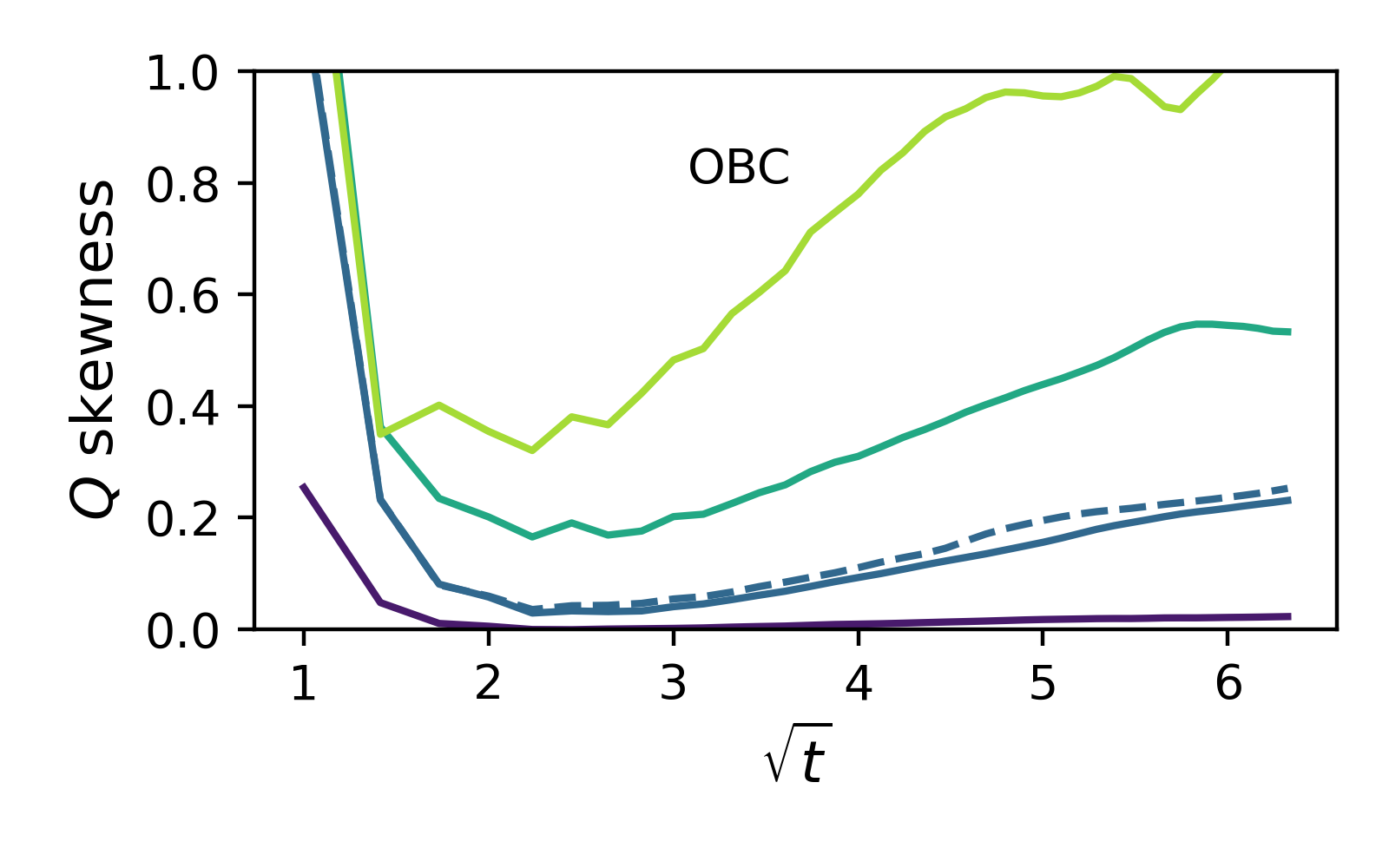}
    \caption{The mean, standard deviation, and skewness of the quantum distribution over $Q$ as a function of time. The top (bottom) panels are for periodic (open) boundaries. As in the main text, the solid lines represent data for $N=30$ spins and a variety of initial polarizations $m$, and the dashed line is at $N=24$ and $m=1/3$. All data is extracted by averaging the distribution over $Q$ over 50 samples.}
    \label{fig:supp_Q_stats}
\end{figure*}

\end{document}